\documentclass[12pt,preprint]{aastex}

\shorttitle{CME-CME Interactions}

\shortauthors{Liu et al.}

\begin{document}

\title{Interactions between Coronal Mass Ejections Viewed in
Coordinated Imaging and In Situ Observations}

\author{Ying D. Liu\altaffilmark{1,2}, Janet G. Luhmann\altaffilmark{1},
Christian M\"{o}stl\altaffilmark{1,3,4}, Juan C.
Martinez-Oliveros\altaffilmark{1}, Stuart D. Bale\altaffilmark{1},
Robert P. Lin\altaffilmark{1,5}, Richard A.
Harrison\altaffilmark{6}, Manuela Temmer\altaffilmark{3}, David F.
Webb\altaffilmark{7}, and Dusan Odstrcil\altaffilmark{8}}

\altaffiltext{1}{Space Sciences Laboratory, University of
California, Berkeley, CA 94720, USA; liuxying@ssl.berkeley.edu}

\altaffiltext{2}{State Key Laboratory of Space Weather, National
Space Science Center, Chinese Academy of Sciences, Beijing, China}

\altaffiltext{3}{Institute of Physics, University of Graz, Austria}

\altaffiltext{4}{Space Research Institute, Austrian Academy of
Sciences, Graz, Austria}

\altaffiltext{5}{School of Space Research, Kyung Hee University,
Yongin, Gyeonggi 446-701, Korea}

\altaffiltext{6}{Space Science and Technology Department, Rutherford
Appleton Laboratory, Didcot, UK}

\altaffiltext{7}{Institute for Scientific Research, Boston College,
Newton, MA 02459, USA}

\altaffiltext{8}{NASA Goddard Space Flight Center, Greenbelt, MD
20771, USA}

\begin{abstract}

The successive coronal mass ejections (CMEs) from 2010 July 30 -
August 1 present us the first opportunity to study CME-CME
interactions with unprecedented heliospheric imaging and in situ
observations from multiple vantage points. We describe two cases of
CME interactions: merging of two CMEs launched close in time and
overtaking of a preceding CME by a shock wave. The first two CMEs on
August 1 interact close to the Sun and form a merged front, which
then overtakes the July 30 CME near 1 AU, as revealed by wide-angle
imaging observations. Connections between imaging observations and
in situ signatures at 1 AU suggest that the merged front is a shock
wave, followed by two ejecta observed at Wind which seem to have
already merged. In situ measurements show that the CME from July 30
is being overtaken by the shock at 1 AU and is significantly
compressed, accelerated and heated. The interaction between the
preceding ejecta and shock also results in variations in the shock
strength and structure on a global scale, as shown by widely
separated in situ measurements from Wind and STEREO B. These results
indicate important implications of CME-CME interactions for shock
propagation, particle acceleration and space weather forecasting.

\end{abstract}

\keywords{shock waves --- solar-terrestrial relations --- solar wind
--- Sun: coronal mass ejections (CMEs)}

\section{Introduction}

Coronal mass ejections (CMEs) are large-scale expulsions of plasma
and magnetic field from the solar atmosphere. One of the most
intriguing questions concerning CMEs is how they interact between
each other during their propagation in interplanetary space. CME-CME
interactions are expected to be a frequent phenomenon near solar
maximum when multiple CMEs can occur within one day, while their
transit time from the Sun to the Earth is typically 4 days.

Interactions between CMEs are of importance for both space weather
studies and basic plasma physics. First, CME-CME interactions can
produce or enhance southward magnetic fields
\citep[e.g.,][]{farrugia04, wang03}, a key factor in geomagnetic
storm generation \citep{dungey61, gosling91}. Second, the
interaction may reveal interesting shock physics in case a shock is
overtaking a CME, including modifications in the shock strength,
particle acceleration and transport. In situ measurements usually
show a depressed plasma $\beta$ within interplanetary CMEs (ICMEs).
The penetrating shock is likely to decay because of the enhanced
Alfv\'{e}n speed in the preceding ejecta. Efficiency in particle
acceleration is expected to change due to modifications in the shock
strength and structure by the preceding ejecta, and particles
accelerated at the shock may be trapped by the closed magnetic
fields within the preceding ejecta or guided along the helical field
lines. (\citet{gopalswamy02} find a close association between CME
interactions and solar energetic particle events, whereas
\citet{richardson03} argue that the association is not statistically
meaningful.) Third, the interaction implies significant energy and
momentum transfer between the interacting CMEs where magnetic
reconnection may take place. Under magnetic reconnection the two
interacting flux systems may finally merge, leading to a phenomenon
called ``CME cannibalism" by \citet{gopalswamy01}. This physical
process would be very complex since CMEs are three-dimensional
large-scale structures.

Studies of CME interactions so far are based on either coronagraph
observations close to the Sun \citep{gopalswamy01} or in situ
measurements near the Earth \citep[e.g.,][]{burlaga02, wang03,
farrugia04}. Connections between imaging observations and in situ
measurements have been lacking. As a result, details of the
interacting process cannot be continuously followed and memory of
the source conditions is lost. With the launch of the Solar
Terrestrial Relations Observatory \citep[STEREO;][]{kaiser08},
coordinated wide-angle imaging and in situ observations of CME
interactions are feasible and can be performed from multiple vantage
points. STEREO is comprised of two spacecraft with one preceding the
Earth (STEREO A) and the other trailing behind (STEREO B). Each of
the STEREO spacecraft carries an identical imaging suite, the Sun
Earth Connection Coronal and Heliospheric Investigation
\citep[SECCHI;][]{howardra08}, which can image a CME from its birth
in the corona all the way to the Earth and beyond. STEREO also has
several sets of in situ instrumentation, which provide in situ
measurements of the magnetic field, energetic particles and the bulk
solar wind plasma \citep{luhmann08, galvin08}. At L1, Wind and ACE
monitor the near-Earth solar wind conditions, thus adding a third
vantage point for in situ measurements.

Around 2010 August 1 the Sun exhibited substantial activities
including filament eruptions, flares and multiple CMEs
\citep{schrijver11, harrison11, juan11, temmer11, mostl12, webb12},
which provides a great opportunity to study CME-CME interactions.
The focus of this Letter is to present the first study of CME
interactions combining wide-angle imaging observations from STEREO
with in situ measurements at 1 AU. The results obtained here are
crucial for understanding CME-CME interactions as well as the
complete picture of CME propagation from the Sun to the Earth.

\section{Observations and Results}

Figure~1 (a) shows the configuration of the planets and spacecraft
in the ecliptic plane on August 1. STEREO A and B are separated by
about $149.9^{\circ}$ in longitude with a distance of 0.96 AU and
1.06 AU from the Sun, respectively. Also shown are propagation
directions of three CMEs of interest in the ecliptic plane
determined from a geometric triangulation technique (see details
below). The first CME (CME1) is launched from the Sun at about 07:30
UT on July 30 with a speed around 540 km s$^{-1}$, while the other
two (CME2 and CME3) are launched at about 02:42 UT and 07:48 UT on
August 1 with speeds around 730 km s$^{-1}$ and 1140 km s$^{-1}$,
respectively. The launch times are estimated by extrapolating the
coronagraph observations of SECCHI back to the solar surface, and
the speeds are obtained from linear fits to CME propagation
distances before CME collisions (see below). A direct impression is
that these CMEs may interact since their propagation directions are
close to each other. CME2 and CME3 are expected to interact at
distances not far from the Sun, while interactions of CME1 with the
other two should take place much further from the Sun because its
launch time is about 2 days earlier.

Figure~1 (b) displays two synoptic views of the CMEs from STEREO A
and B. Only shown are data from the outer coronagraph (COR2) and the
heliospheric imagers (HI1 and HI2) of SECCHI. COR2 has a field of
view (FOV) of 0.7$^{\circ}$ - 4$^{\circ}$ around the Sun. HI1 has a
20$^{\circ}$ square FOV centered at 14$^{\circ}$ elongation from the
center of the Sun while HI2 has a 70$^{\circ}$ FOV centered at
53.7$^{\circ}$. HI1 and HI2 can observe CMEs to the vicinity of the
Earth and beyond by using sufficient baffling to eliminate stray
light \citep{harrison08, eyles09}. Note a data gap for STEREO B from
10 UT of August 1 to 04 UT of August 2. CME1 is largely north of the
ecliptic plane but rotates and expands toward the plane. Spacecraft
in the ecliptic plane will likely encounter its flank. CME2 and CME3
are propagating more along the ecliptic plane. CME3 is fast and
energetic (1140 km s$^{-1}$), and it seems to overtake CME2 in the
FOV of HI1. A merged front is formed from the interaction between
CME2 and CME3. Connections with in situ signatures at 1 AU suggest
that this merged front is a shock wave. CME1 is then overtaken by
the merged front at elongations close to the Earth, as shown in HI2
of STEREO A.

Figure~2 shows the time-elongation maps, which are produced by
stacking the running difference intensities of COR2, HI1 and HI2
within a slit around the ecliptic plane \citep[e.g.,][]{sheeley08,
davies09, liu10a}. At least 4 CMEs occurred on August 1, whereas
between July 30 and August 1 the Sun was relatively quiet. (There
appears a weak feature ahead of CME2 in COR2 of STEREO A, which is
produced by small plasma flows along a coronal streamer.)
\citet{harrison11} provide an overview of the CMEs on August 1. As
can be seen from the maps, tracks from CME2 and CME3 intersect,
indicative of an interaction between these two events. Some of the
tracks merge into a single one in the FOV of HI1, and later a
bifurcation is observed. Presumably, the leading bifurcated feature
is the shock wave, and the trailing one is another structure visible
to STEREO A after the CME-CME interaction (also see
\citet{harrison11} for more discussions on the bifurcated
structures). There is a gap in STEREO B observations during the time
of this interaction, but the tracks from both spacecraft before and
after the data gap can be used to calculate the propagation
direction and distance of corresponding features with a geometric
triangulation method. Also shown in Figure~2 are fits to the leading
tracks of CME1 in HI1 and HI2 by assuming a kinematic model with a
constant speed and propagation direction \citep[e.g.,][]{sheeley08}.
The fit is a rough representation of the tracks given its
assumptions. The propagation direction determined from the track
fitting is about $28^{\circ}$ and $39^{\circ}$ east of the Sun-Earth
line for STEREO A and B, respectively. The extrapolated fit curves
suggest that the interaction between CME1 and the merged front of
CME2 and CME3 would probably occur around 1 AU.

The elongation angles along the tracks can be converted to radial
distance and propagation direction using a geometric triangulation
method developed by \citet{liu10a, liu10b}. The technique has had
success in tracking CMEs and connecting imaging observations with in
situ signatures for various CMEs and spacecraft longitudinal
separations \citep{liu10a, liu10b, liu11, mostl10}. We apply the
technique to the leading features of the three CMEs as well as the
merged front of CME2 and CME3. The resulting CME kinematics in the
ecliptic plane are displayed in Figure~3. The propagation direction
is converted to an angle with respect to the Sun-Earth line. If the
angle is positive (negative), the CME feature would be propagating
west (east) of the Sun-Earth line. The propagation angles show a
variation with time, with an average value of $-23^{\circ}$ for CME2
and $-19^{\circ}$ for CME3 before their collision. The speed
obtained from the linear fit to the distances is about 732 km
s$^{-1}$ and 1138 km s$^{-1}$ for CME2 and CME3, respectively. These
propagation angles and speeds are consistent with estimates from
radio type II bursts \citep{juan11} and other various methods
\citep{temmer11}. The merging between CME2 and CME3 is likely
complete around 17 UT on August 1 at a distance of about 55 solar
radii from the Sun, estimated by extrapolating the fits to the point
where they intersect.

The kinematics of the merged front of CME2 and CME3 are shown in the
right panels of Figure~3. The average propagation angle is about
$-22^{\circ}$, which has not changed much compared with those before
the collision. The merged front can be tracked out to about 165
solar radii or 0.75 AU (without projection). The speed obtained from
the linear fit to the distances is about 621 km s$^{-1}$, smaller
than the speeds of both CMEs before their interaction. Note that
this is the speed well after the interaction. While the main
deceleration of the overtaking CME (CME3) is due to the interaction,
the solar wind drag may also contribute to the slowdown
\citep{temmer11}. The predicted arrival time of the merged front at
the Earth, estimated from the linear fit, is about 17:36 UT on
August 3. Also indicated in Figure~3 is the average propagation
angle of CME1. Application of the triangulation technique to this
CME gives propagation angles that show a transition from
$-12^{\circ}$ to $-40^{\circ}$ with an average value of about
$-30^{\circ}$ (not included in this Letter). This average
propagation angle is consistent with the estimates from track
fitting ($-28^{\circ}$ and $-39^{\circ}$ for STEREO A and B,
respectively). The merged front of CME2 and CME3 would probably
interact with CME1 given their similar propagation angles and
large-scale structures.

Figure~4 shows the in situ measurements at Wind. Three ICMEs can be
identified from the Wind data between August 3 - 5. A strong forward
shock passed Wind at 17:04 UT on August 3. The predicted arrival
time (17:36 UT) of the merged front of CME2 and CME3 is coincident
with the shock passage at Wind, which suggests that the merged front
in white-light images is the shock. The shock is overtaking a
preceding ICME (ICME1) at 1 AU. The trailing boundary of ICME1 is
mainly determined from the low proton $\beta$, while the leading
boundary can be identified from the smooth, slightly enhanced
magnetic field and depressed proton temperature (compared with the
expected temperature) in addition to the proton $\beta$. Although
shock compression can enhance the plasma density, temperature and
magnetic field, the plasma $\beta$ may not change much. This is why
we use the proton $\beta$ to determine the trailing boundary of
ICME1. ICME1 is also observed at STEREO B (see Figure~5), for which
the rotation of the magnetic field components and proton $\beta$ are
used together to determine the boundaries. The predicted arrival
time of CME1 is about 17:26 UT and 21:56 UT on August 2 at Wind and
STEREO B, respectively, which has a close timing with the in situ
measurements of ICME1. This suggests that ICME1 is the corresponding
structure of CME1 at 1 AU. The fact that the shock is overtaking
ICME1 is consistent with the imaging observations that indicate an
interaction between CME1 and the merged front of CME2 and CME3 close
to 1 AU. The predicted speed (about 621 km s$^{-1}$) of the shock is
consistent with but slightly larger than observed at Wind.
Presumably the shock is slowed down by the interaction.

The shock significantly accelerates, compresses and heats the
preceding ejecta, as can be seen from Figure~4. First, the trailing
edge of ICME1 is traveling faster than the leading edge (with a
speed difference of about 170 km s$^{-1}$ at Wind), as the trailing
edge is downstream of the shock while the leading edge upstream. As
a result, the radial width of ICME1 is decreasing. After the shock
has gone through the whole ejecta, the speed of every part of the
ejecta would be increased, so ICME1 is essentially accelerated by
the shock. The plasma density and magnetic field within the ejecta
would also be enhanced because of the shock compression. Second, the
proton temperature within ICME1 is low (only about 20000 K upstream
of the shock at Wind), which results in a large sound Mach number of
the shock ($M_s \sim 15$). The observed temperature increases
significantly across the shock (by a factor of about 15), which is
expected as the heating by a hydrodynamic shock is proportional to
$M_{s}^{2}$ although the magnetic field may reduce the heating
somehow. MHD simulations of a shock overtaking a preceding ejecta
seem to give similar results \citep[e.g.,][]{vandas97, schmidt04,
lugaz05, xiong06}.

A similar scenario is observed at STEREO B (see Figure~5), but the
measurements at this additional point indicate a significant
distortion in both the shock strength and structure. First, the
Alfv\'{e}n speed upstream the shock at STEREO B is about 150 km
s$^{-1}$, much larger than that upstream the shock at Wind (about 40
km s$^{-1}$). This reduces the shock Alfv\'{e}n Mach number, $M_A
\sim 1.5$ at STEREO B compared with about 6 at Wind. Consequently
the shock strength is decreased with a density compression ratio of
about 2 and magnetic field compression ratio of about 1.8 at STEREO
B (compared with 3 for the density and 4 for the field at Wind).
Efficiency in particle acceleration is expected to decrease too.
Second, the shock arrives at STEREO B around 04:55 UT on August 3,
about 12 hours earlier than at Wind, although STEREO B is further
away from the Sun (1.06 AU). This indicates a non-spherical
structure of the shock. A possible explanation is that CME1, whose
propagation direction is east of the shock, may have removed some of
the solar wind plasma ahead of the shock. The eastern flank of the
shock is thus expected to move faster as it is propagating into a
less dense medium. A closer look at Figures~4 and 5 also reveals a
pronounced northward magnetic field component downstream the shock
at Wind while a large southward field component behind the shock at
STEREO B, presumably produced by the interaction between the shock
and preceding ejecta.

In addition to the interaction of a shock with a preceding ejecta,
we also observe two ICMEs (ICME2 and ICME3) behind the shock at Wind
that seem to have already merged. Connections between the imaging
observations and in situ measurements suggest that ICME2 and ICME3
correspond to CME2 and CME3, respectively, although other
possibilities cannot be completely excluded. ICME2 is mainly
identified from the rotation of the magnetic field and depressed
proton $\beta$, while the identification of ICME3 is
straightforward. A region with relatively enhanced proton density,
temperature and $\beta$ is observed between ICME2 and ICME3.
Presumably this is the interface of the CME-CME interaction. The
magnetic field polarity in this interaction region is opposite to
those inside the two ejecta, so magnetic reconnection may have
occurred. The interval of ICME2 is very short, and its temperature
seems enhanced compared with a typical ICME at 1 AU. A possible
explanation is that the shock driven by ICME3, which is probably
shock2, has already passed through ICME2 at 1 AU. This explanation
is consistent with imaging observations. Compression by ICME3 from
behind may also contribute to the heating and field enhancement in
addition to shock compression. In situ measurements at STEREO B also
indicate an ejecta-like structure behind the shock, whose
identification is mainly based on the relatively enhanced magnetic
field. Its plasma and magnetic field structure, however, is very
complicated and does not qualify for a typical ICME. One ICME may
miss STEREO B, or the merging of the two ICMEs is such that typical
ICME signatures are no longer recognizable. The readers are directed
to \citet{mostl12} for more discussions on the in situ signatures.

\section{Summary}

We have investigated CME-CME interactions, combining imaging
observations with in situ measurements from multiple vantage points.
With the advantage of having wide-angle imaging observations, we are
able to follow how the patterns of interacting CMEs evolve with time
and how the interaction features in images connect with in situ
signatures. Two CMEs (CME2 and CME3) from 2010 August 1 merged
around 55 solar radii from the Sun into a broad wave with enhanced
brightness. Connections with in situ signatures suggest that the
merged front is a shock, followed by two ejecta observed at Wind
which seem to have already merged. The shock, which is probably
driven by CME3, may have passed through CME2 well before 1 AU and is
propagating into the CME from July 30 (CME1) near 1 AU. In situ
measurements at 1 AU show that the preceding ejecta is significantly
compressed, accelerated and heated by the overtaking shock. The
interaction also modifies the shock strength and structure on a
global scale as indicated by additional measurements at STEREO B.

\acknowledgments The research was supported by the STEREO project
under grant NAS5-03131. C. M. is supported by a Marie Curie
International Outgoing Fellowship. M. T. acknowledges the Austrian
Science Fund FWF V195-N16.

\clearpage

\begin{figure}
\epsscale{1.0} \plotone{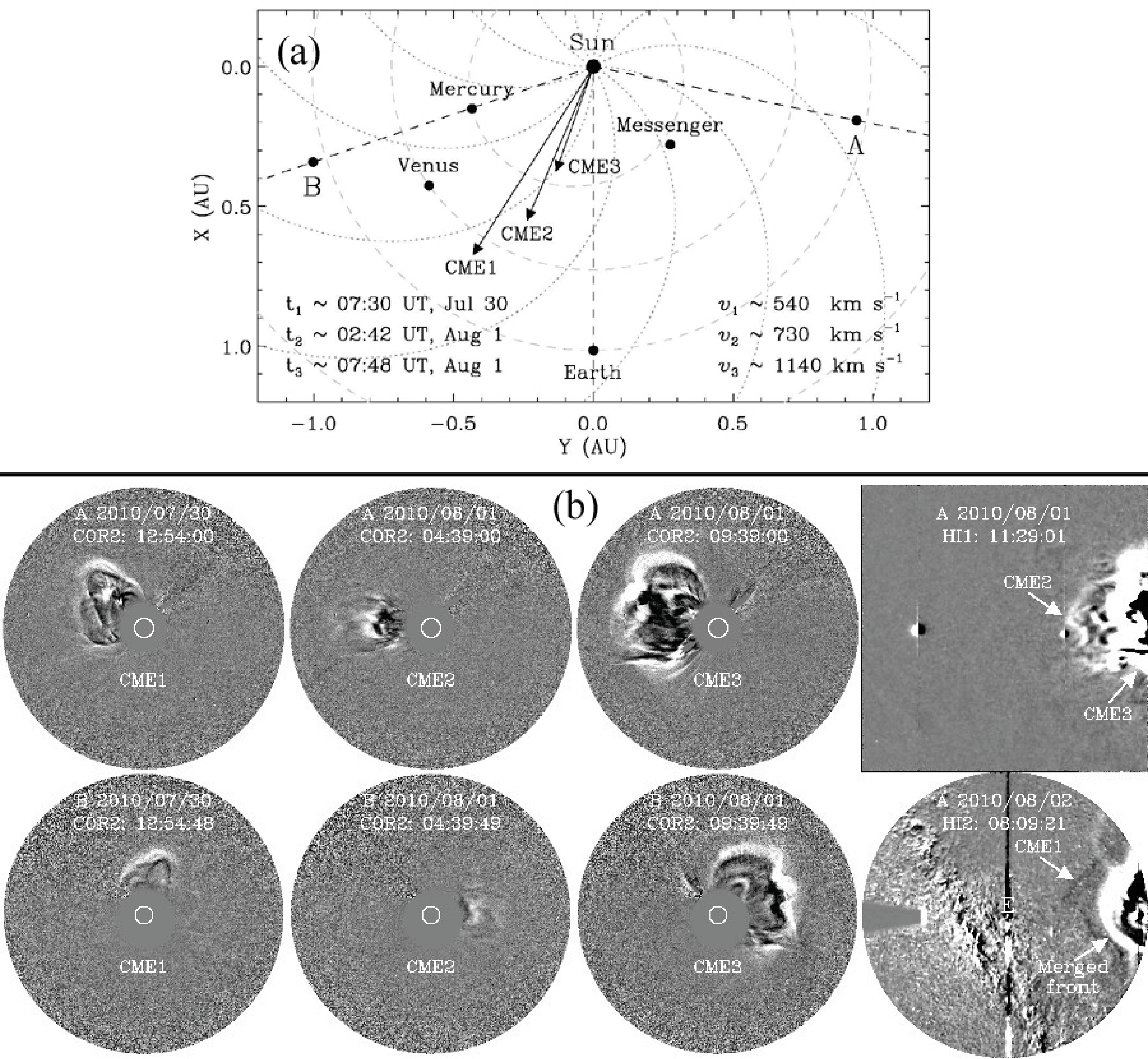} \caption{(a) Positions of the
spacecraft and planets in the ecliptic plane on 2010 August 1. The
gray dashed curves indicate the planetary orbits, and the dotted
lines show Parker spiral magnetic fields. The arrows mark the
propagation directions of the CMEs of interest obtained from a
triangulation technique. The estimated CME speeds and launch times
on the Sun are also given. (b) CME evolution observed by STEREO. The
left three columns display COR2 images of the three CMEs viewed from
STEREO A (upper) and B (lower) near simultaneously. The right column
shows images from HI1 and HI2 of STEREO A, which indicate the
scenario of CME-CME interactions. The position of the Earth is
labeled as E. (Animations of this figure are available in the online
journal.)}
\end{figure}

\clearpage

\begin{figure}
\epsscale{0.8} \plotone{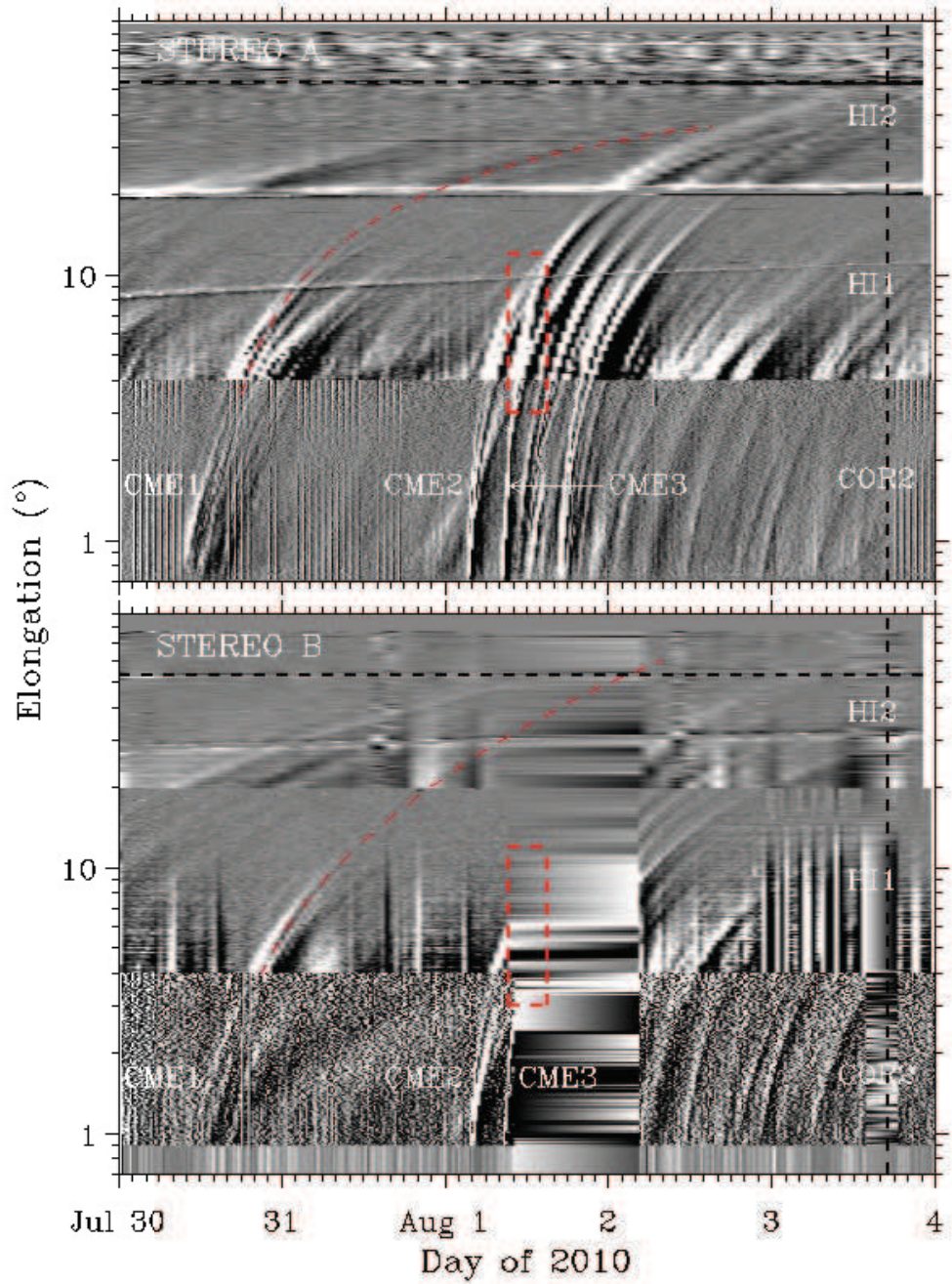} \caption{Time-elongation maps
constructed from running difference images of COR2, HI1 and HI2
along the ecliptic plane for STEREO A (upper) and B (lower). Tracks
associated with the three CMEs are indicated. The rectangular box
marks the times and elongation angles of the interaction between
CME2 and CME3. The vertical dashed line indicates the observed
arrival time of a shock at the Earth, and the horizontal dashed line
denotes the elongation angle of the Earth. The red curve shows the
fit to the leading tracks of CME1 in HI1 and HI2.}
\end{figure}

\clearpage

\begin{figure}
\centerline{\includegraphics[width=20pc]{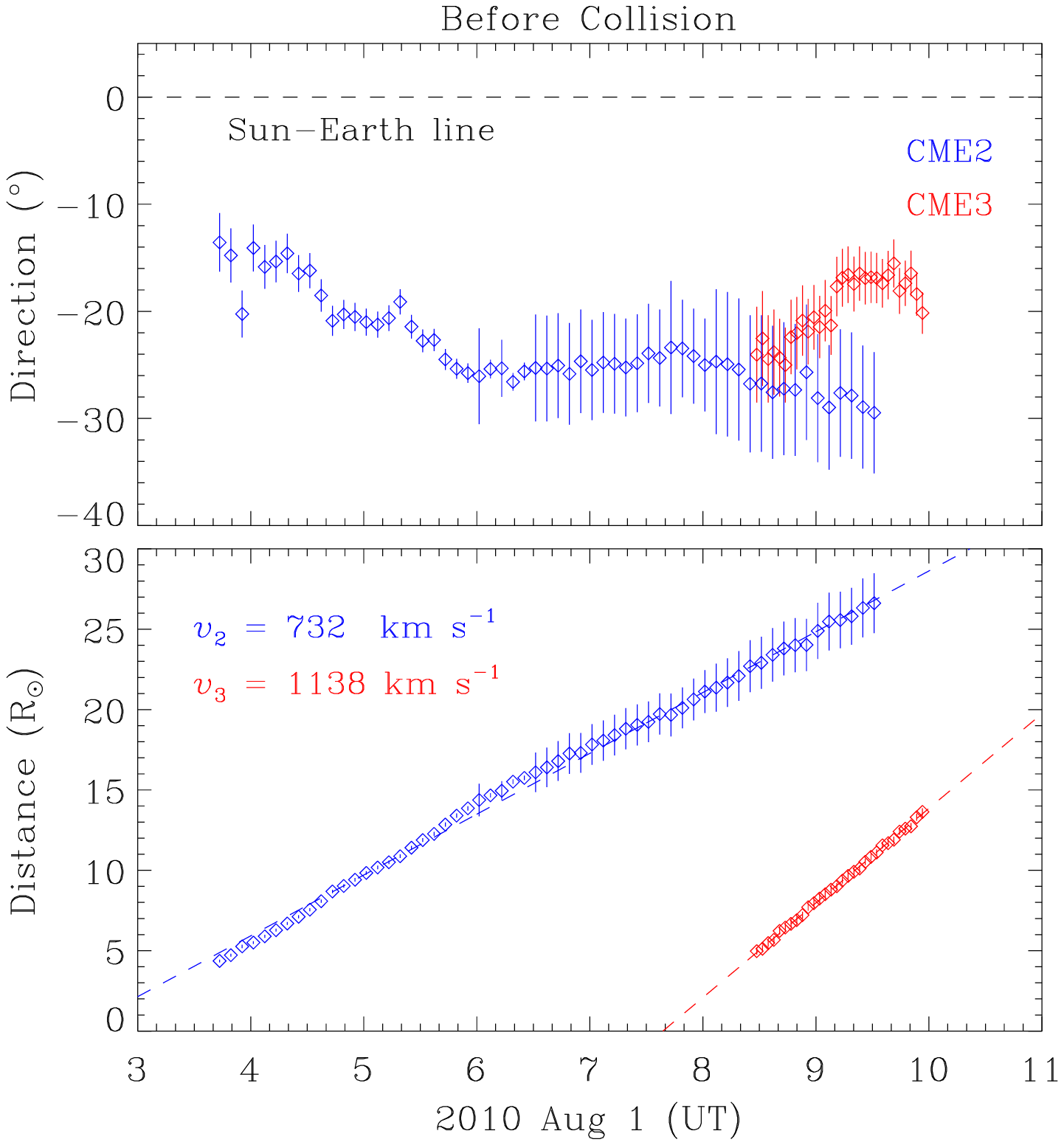}
\includegraphics[width=20pc]{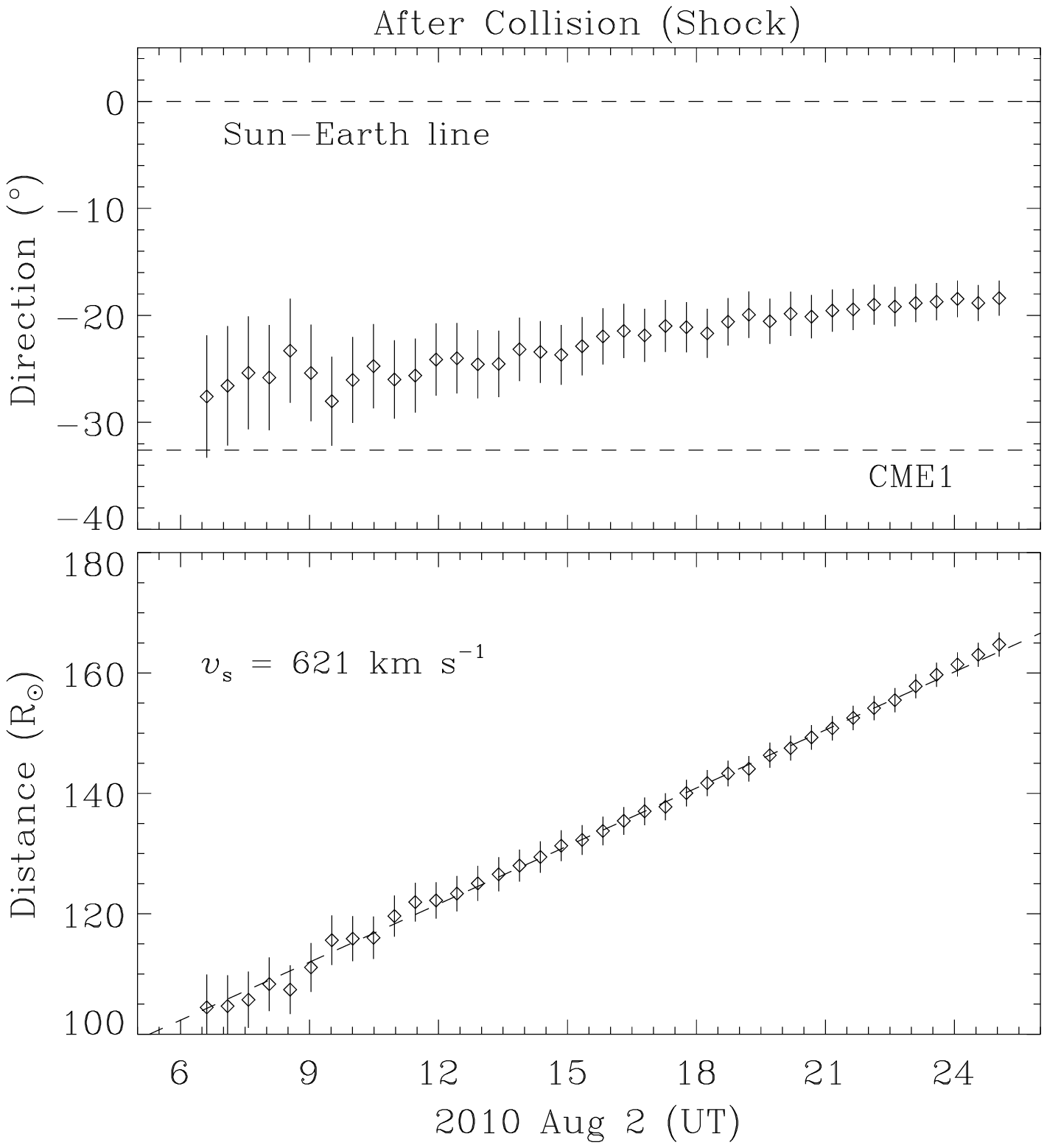}}
\caption{Propagation direction and radial distance derived from
geometric triangulation before (left) and after (right) collision
between CME2 and CME3. The lower dashed line in the top right panel
indicates the propagation angle of CME1. Also shown are linear fits
to the distances, together with the speeds obtained from the fits.}
\end{figure}

\clearpage

\begin{figure}
\epsscale{0.7} \plotone{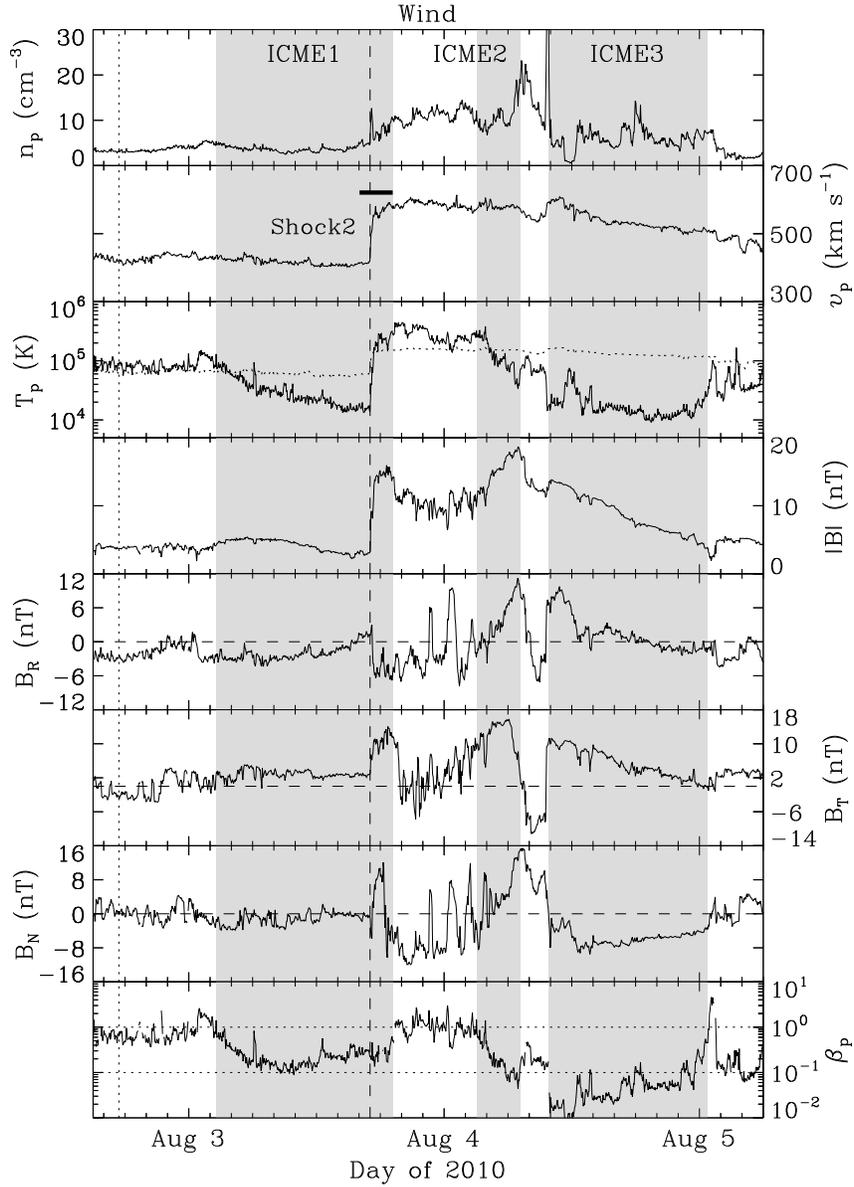} \caption{Solar wind plasma and
magnetic field parameters observed at Wind. From top to bottom, the
panels show the proton density, bulk speed, proton temperature,
magnetic field strength and components, and proton $\beta$,
respectively. The dotted curve in the third panel denotes the
expected proton temperature from the observed speed. The shaded
regions show the ICME intervals, and the vertical dashed line
indicates the associated shock. The horizontal line in the second
panel marks the predicted arrival time (with uncertainties) and
speed of the merged front of CME2 and CME3 at the spacecraft. The
predicted arrival time of CME1 is indicated by the vertical dotted
line.}
\end{figure}

\clearpage

\begin{figure}
\epsscale{0.7} \plotone{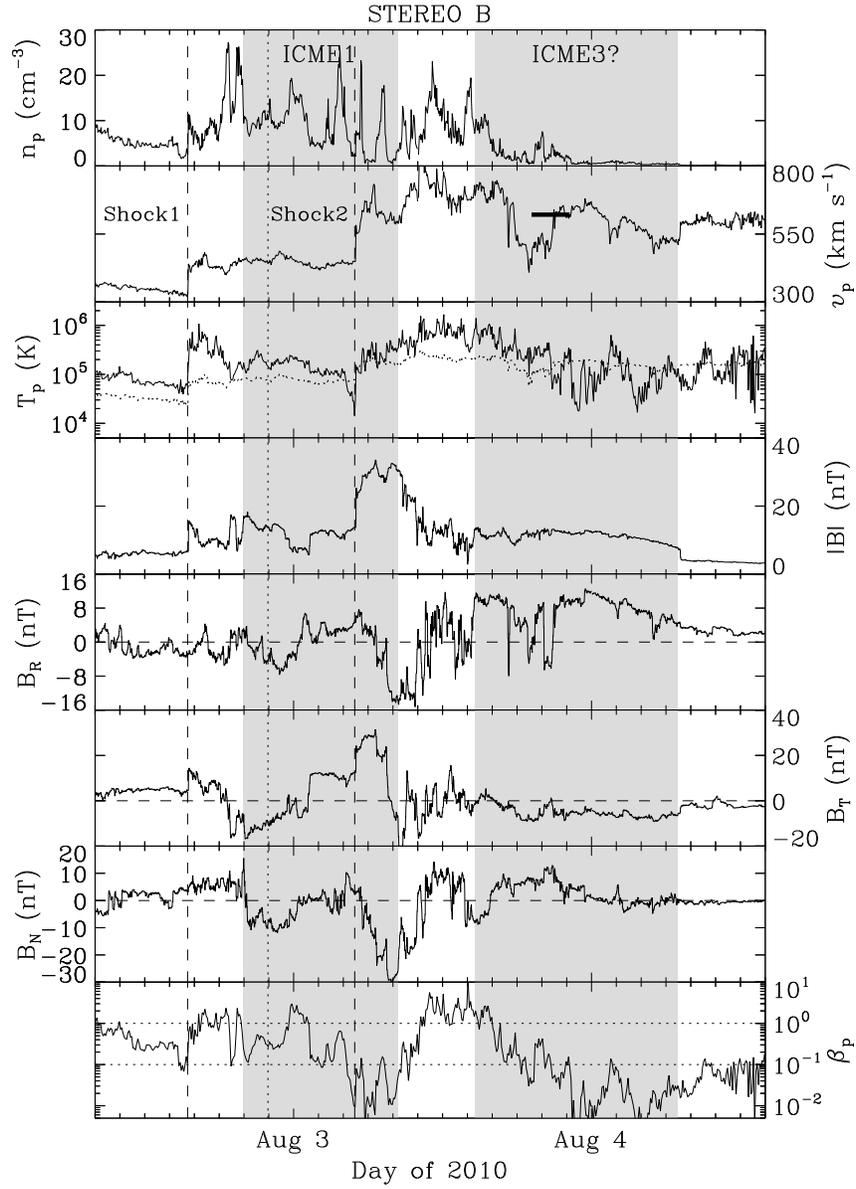} \caption{Similar format to Figure~4,
but for the measurements at STEREO B. ICME1 has a forward shock at
STEREO B, while only a discontinuity is observed at Wind.}
\end{figure}

\end{document}